\let\Xdocument\document
\documentclass{iau}
\let\document\Xdocument

\usepackage{amsmath}
\usepackage{graphicx}
\usepackage{multirow}
\usepackage[flushleft]{threeparttable}

\begin{document}
\lefttitle{K. Kurbah et al.}
\righttitle{Multiphase study of Cepheids in the MCs}

\jnlPage{1}{8}
\jnlDoiYr{2023}
\doival{10.1017/xxxxx}

\aopheadtitle{Proceedings IAU Symposium 376}
\editors{Richard de Grijs, Patricia Whitelock and M\'arcio Catelan, eds.}

\title{A multiphase study of classical Cepheids in the Magellanic Clouds- Models and Observations}

\author{Kerdaris Kurbah$^1$, Shashi M. Kanbur $^2$, Sukanta Deb $^1$, Susmita Das $^3$, Mami Deka $^1$,  Anupam Bhardwaj$^4$, Hugh Riley Randal $^2$, Selim Kalici $^2$}
\affiliation{$^1$ Department of Physics, Cotton University, Guwahati, Assam, 781001, India\\
$^2$ Department of Physics, State University of New York, Oswego, NY 23126, USA \\
$^3$ Konkoly Observatory, Research Centre for Astronomy and Earth Sciences, Eötvös Loránd Research Network (ELKH), Konkoly-Thege Miklós út 15-17, H-1121, Budapest, Hungary\\
$^4$ INAF-Osservatorio Astronomico di Capodimonte, Via Moiariello 16, I-801301, Napoli, Italy}

\begin{abstract}
This work presents the study of multiphase relations of classical Cepheids in the Magellanic Clouds for short periods ($\log{P}<1$) and long periods ($\log{P}>1$). From the analysis, it has been found that the multiphase relations obtained using the models as well as observations are highly dynamic with pulsational phase. The multiphase relations for short and long periods are found to display contrasting behaviour for both LMC and SMC. It has been observed that the multiphase relations obtained using the models agree better with the observations in the PC plane in most phases in comparison to the PL plane. Multiphase relations obtained using the models display a clear distinction among different convection sets in most phases. Comparison of models and observations in the multiphase plane is one way to test the models with the observations and to constrain the theory of stellar pulsation. 
\end{abstract}

\maketitle

\section{Introduction}
 Theoretical pulsation properties of Cepheid variables have been investigated by Bono and his group in the 1990s \citep{1998ApJ...497L..43B,1999ApJ...512..711B, 2000A&A...360..245B, 2000A&A...359.1059C,2000A&A...354..610C}. The comparison between the observed and theoretical period-color (PC), amplitude-color (AC) and period-luminosity (PL) relations for Cepheids, RR Lyraes, and $\delta$ Scuti stars in the  Galaxy and the Magellanic Clouds in the literature were mostly done at maximum, minimum, and mean light \citep{2004MNRAS.354..212K, 2006MNRAS.369..705K, 2007MNRAS.380..819K, 2020MNRAS.493...29D, 2020ApJS..247..30D, 2022MNRAS.517.2251D, 2022ApJS..262...25D}. However, the relation at mean light is the average of the relations at all phases, hence, investigating the Cepheid PL/PC relations at other phases of the stars provides a better understanding of the pulsational properties of the stars. A multiphase study of Cepheid PL/PC/AC relations has been done by \citet{2006MNRAS.369..723N} and the comparative study of the same with the models has been carried out by \citet{2010MNRAS.408..695K}. In this study, we investigate the theoretical PL/PC relations of Cepheids at multiple phases during a pulsation cycle, using the radial stellar pulsation (RSP) code in  Modules for Experiments in Stellar Astrophysics (\textsc{mesa}, \citet{2011ApJS..192....3P, 2013ApJS..208....4P, 2015ApJS..220...15P, 2018ApJS..234...34P,2019ApJS..243...10P}) and the same for the observations using OGLE-IV database.

\section{Methodology}

Theoretical light curves of Cepheids are generated using \textsc{mesa-rsp} \citep{2008AcA....58..193S,2019ApJS..243...10P} in all four sets of convection parameters A, B, C and D. A brief explanation of the parameters in each convection set is outlined in \citet{2019ApJS..243...10P}. The chemical composition appropriate for the LMC ($Z=0.008, X=0.742$) and SMC ($Z=0.004, X=0.746$) are adopted from \citet{1998ApJ...497L..43B}. For each chemical composition and mass in the range  $3.6~M_{\odot}-6.8~M_{\odot}$, the luminosity is calculated using the rotation-averaged mass-luminosity (ML) relation of \citet{2014A&A...564A.100A}. The models cover a wide range of effective temperatures between $\sim 4000$ K to $8000$ K. It is to be noted that the Cepheid light curves obtained using the models pulsate in the fundamental mode (FU) and fulfilled the condition of full-amplitude and stable pulsation state where the amplitude of radius variation ($\Delta R$), the pulsation period computed on a cycle to cycle basis ($P$), and the fractional growth of the kinetic energy per pulsation period ($\Gamma$) vary by less than $0.01$ in the last $100$ pulsation cycles of the total pulsation cycles computed \citep{2021MNRAS.501..875D}. The period range of the light curves generated using the models is between $3 < {P} (\rm d) < 32$.

Light curves for FU Cepheids in the optical bands ($VI$) in the Magellanic Clouds (MCs) are obtained from the OGLE-IV catalogue \citep{2015AcA....65..297S}. Stars listed in 'remarks.text' file provided by OGLE-IV are rejected. Light curves consisting of more than $30$ data points and within period range of $3 < {P} (\rm d) < 32$ are selected from both the bands leaving $1530$ and $703$ stars in the LMC and SMC, respectively. The apparent magnitudes are corrected for interstellar extinction using \citet{2021ApJS..252...23S} reddening map, the extinction values are adopted using the reddening law of \citet{1989ApJ...345..245C}. Furthermore, the same period range of $3<P(\rm d)<32$ for models and observations is considered throughout the analysis. 

The light curves obtained from the observations and the models are fitted using a cosine Fourier decomposition \citep[and reference therein]{2009A&A...507.1729D} 
\begin{equation}
m(\Phi)= A_{0} + \sum_{i=1}^{N} A_{i}\cos\left(2\pi i\Phi + \phi_{i}\right)\label{equa:e1}
\end{equation}
$\Phi$ is the pulsation phase ranges from $0$ to $1$ given by 
\begin{equation}
\Phi = \frac{(t-t_{0})}{P} - {\rm Int} \left[\frac{(t-t_{0})}{P}\right],
\label{equa:n02}
\end{equation}
where $t$ is the time of observation in days, $t_{0}$ is the epoch of maximum light and $P$ denotes the period in days. 

To investigate the behaviour of the PL/PC relation as a function of phase, equation~(\ref{equa:e1}) is interpolated in $\Phi$ from $0$ to $1$ in steps of $0.02$ on the Fourier-fitted data points. The magnitudes corresponding to the $50$ pulsation phases were extracted from the interpolated light curves for both the models and observations. The multiphase study was carried out for short periods ($\log{P}<1$) and long periods ($\log{P}>1.0$) in both the LMC and SMC. The number of light curves used in the analysis for both LMC and SMC are listed in Table~\ref{tab:model_number_lmcsmc}.

\section{Results}
\begin{figure}
\includegraphics[scale=0.5]{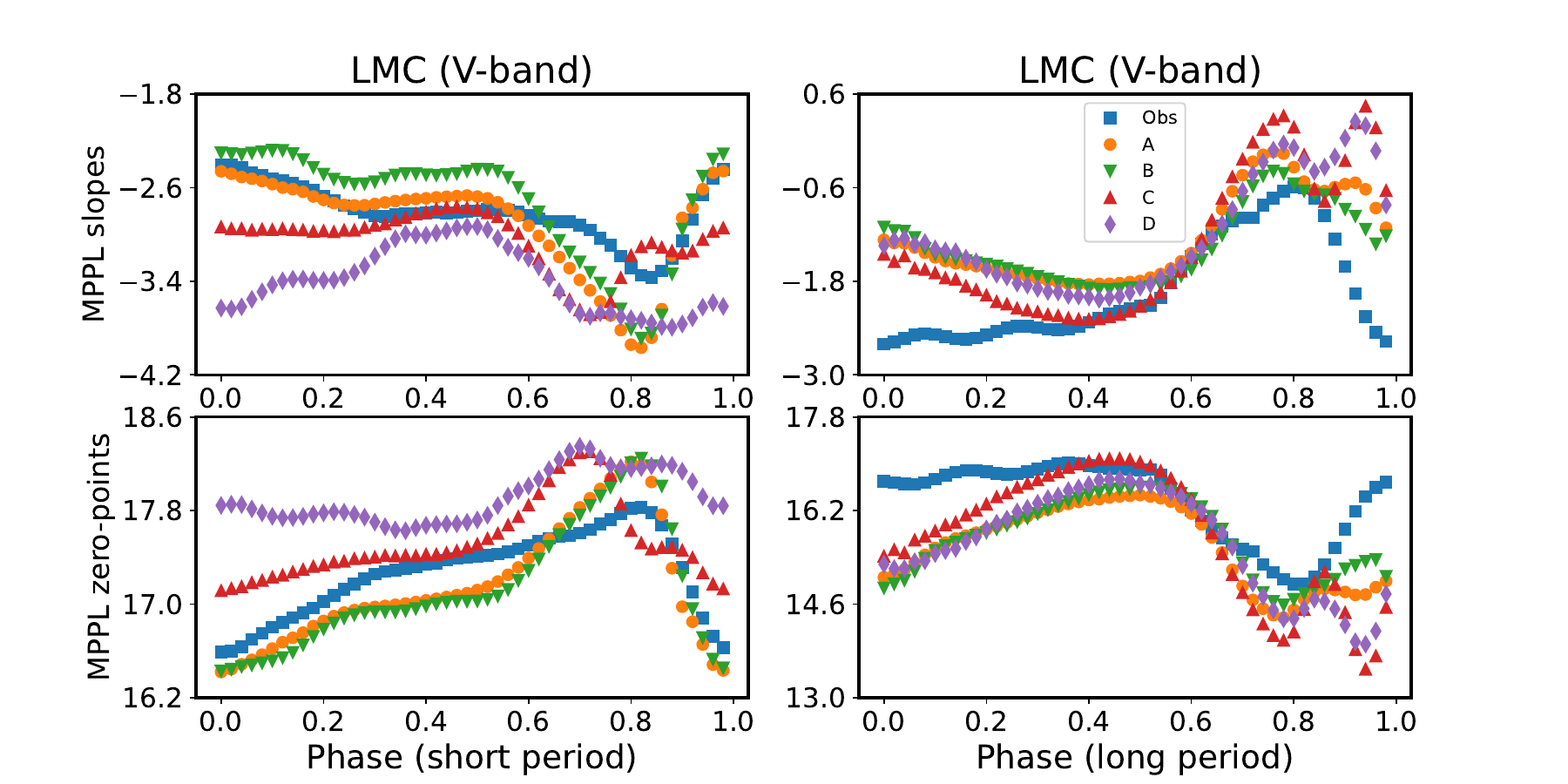}
\caption{Plots of the MPPL relations obtained from Cepheid models and observations for the LMC in $V$-band. Orange, green, red and purple denote the relations obtained using convection sets A, B, C and D, respectively. Blue represent observed relations. Left/right panels represent the relations obtained using short/long period light curves.}
\label{mppl_lmc}
\end{figure}

\begin{figure}
\includegraphics[scale=0.5]{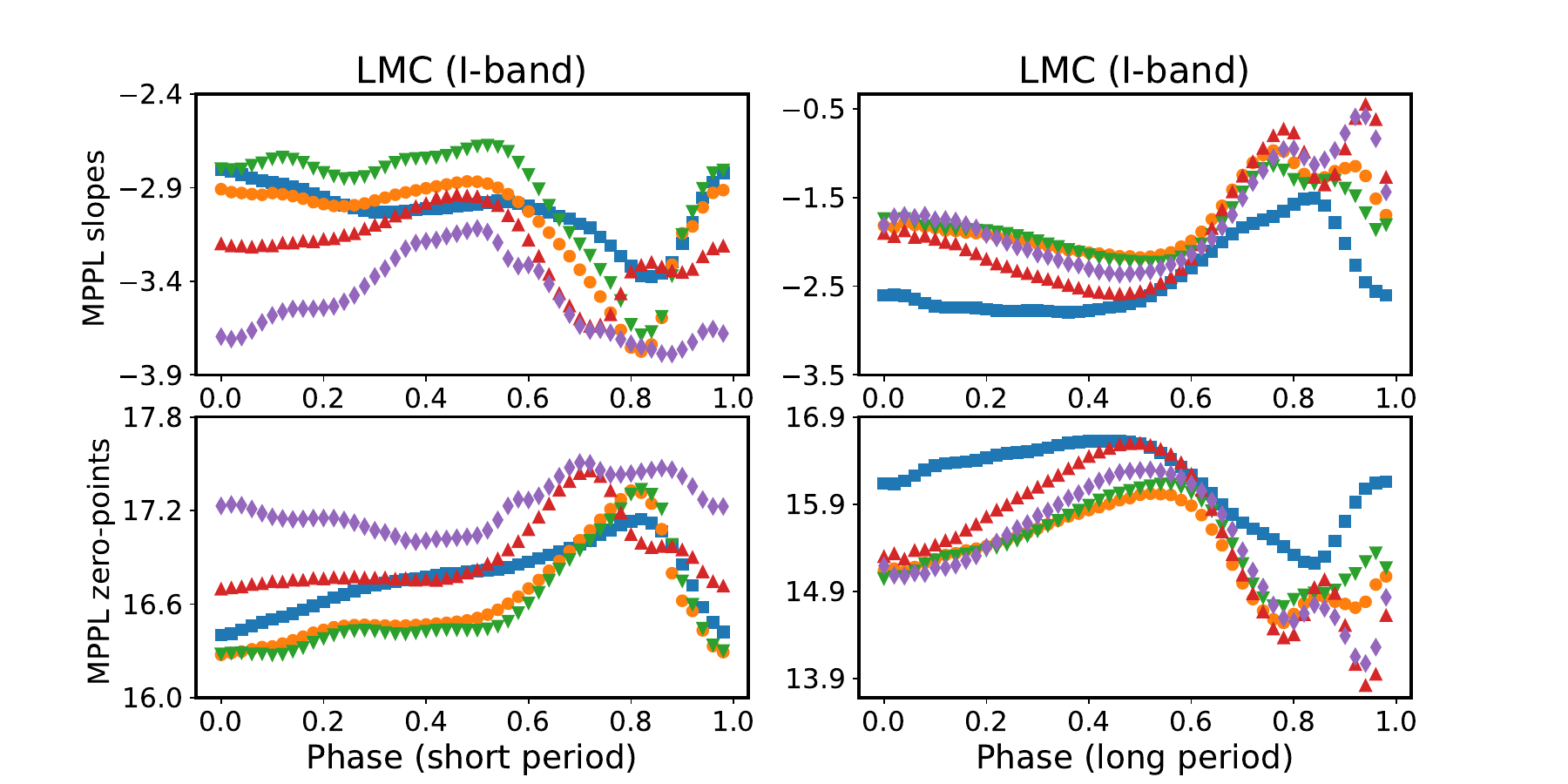}
\caption{Same as Figure~\ref{mppl_lmc} but for LMC in $I$-band.}
\label{mppl_lmci}
\end{figure}

\begin{figure}
\includegraphics[scale=0.5]{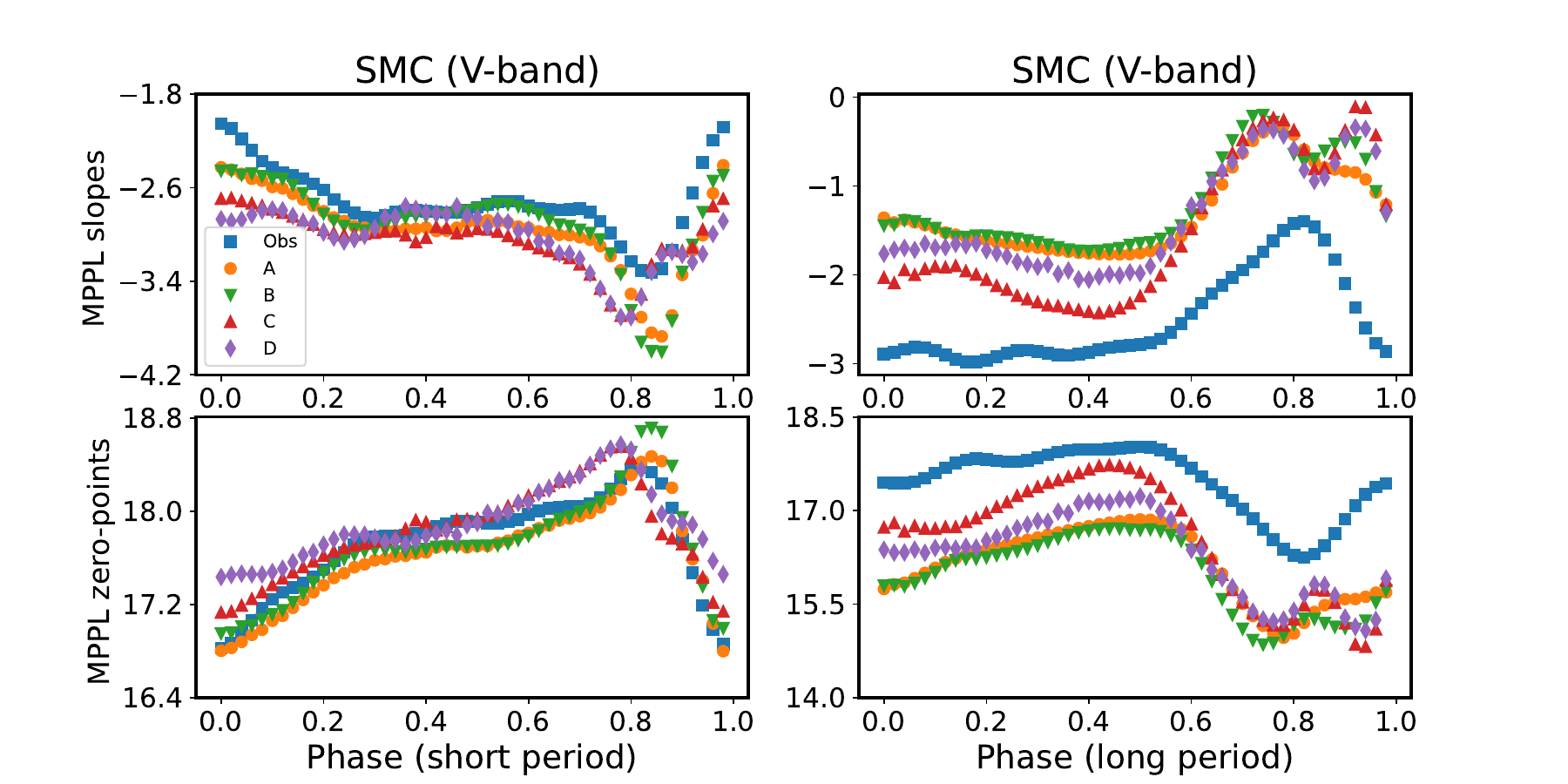}
\caption{Same as Figure~\ref{mppl_lmc} but for SMC.}
\label{mppl_smc}
\end{figure}

\begin{figure}
\includegraphics[scale=0.5]{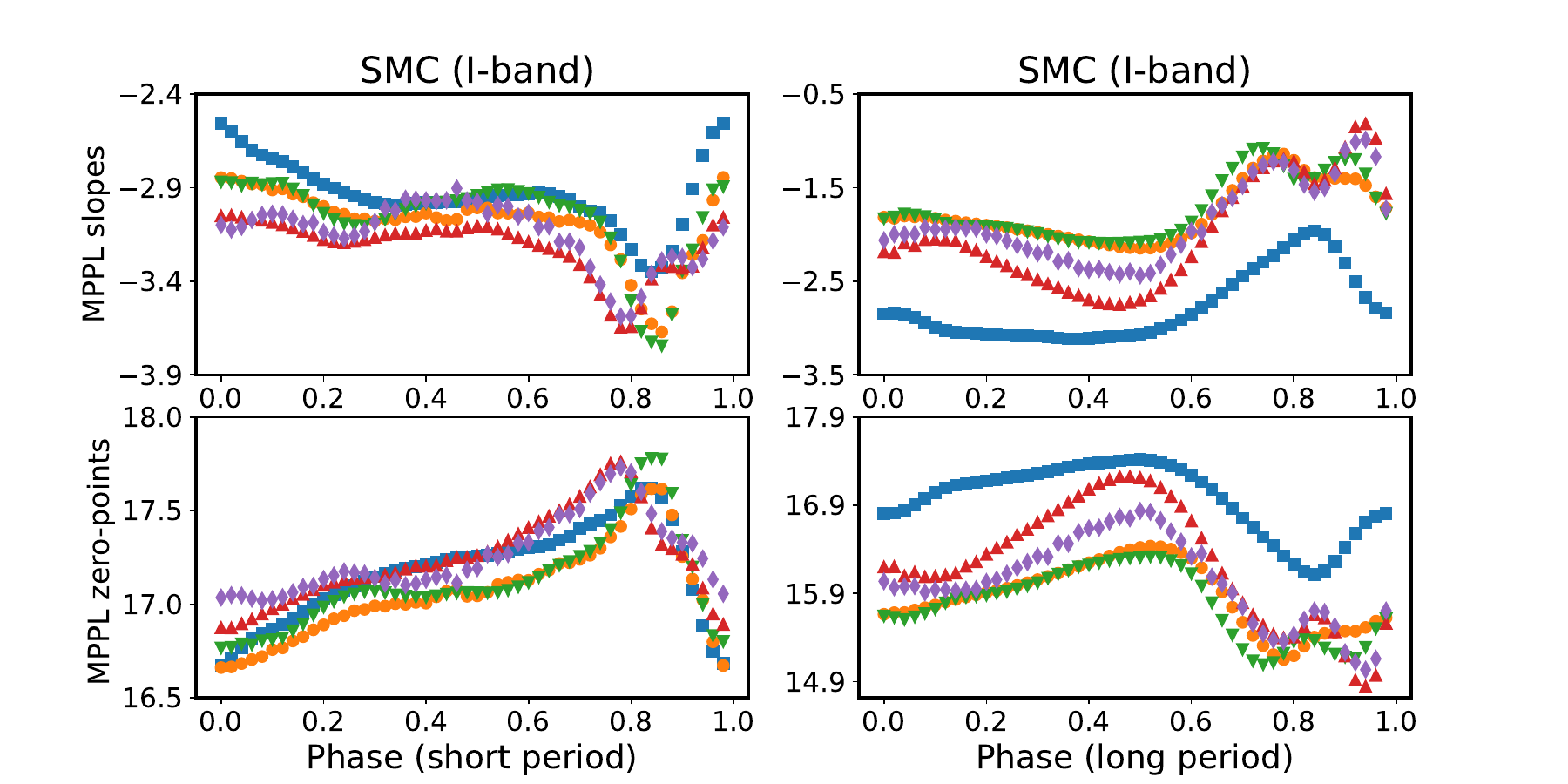}
\caption{Same as Figure~\ref{mppl_lmc} but for SMC in $I$-band.}
\label{mppl_smci}
\end{figure}

\begin{figure}
\includegraphics[scale=0.5]{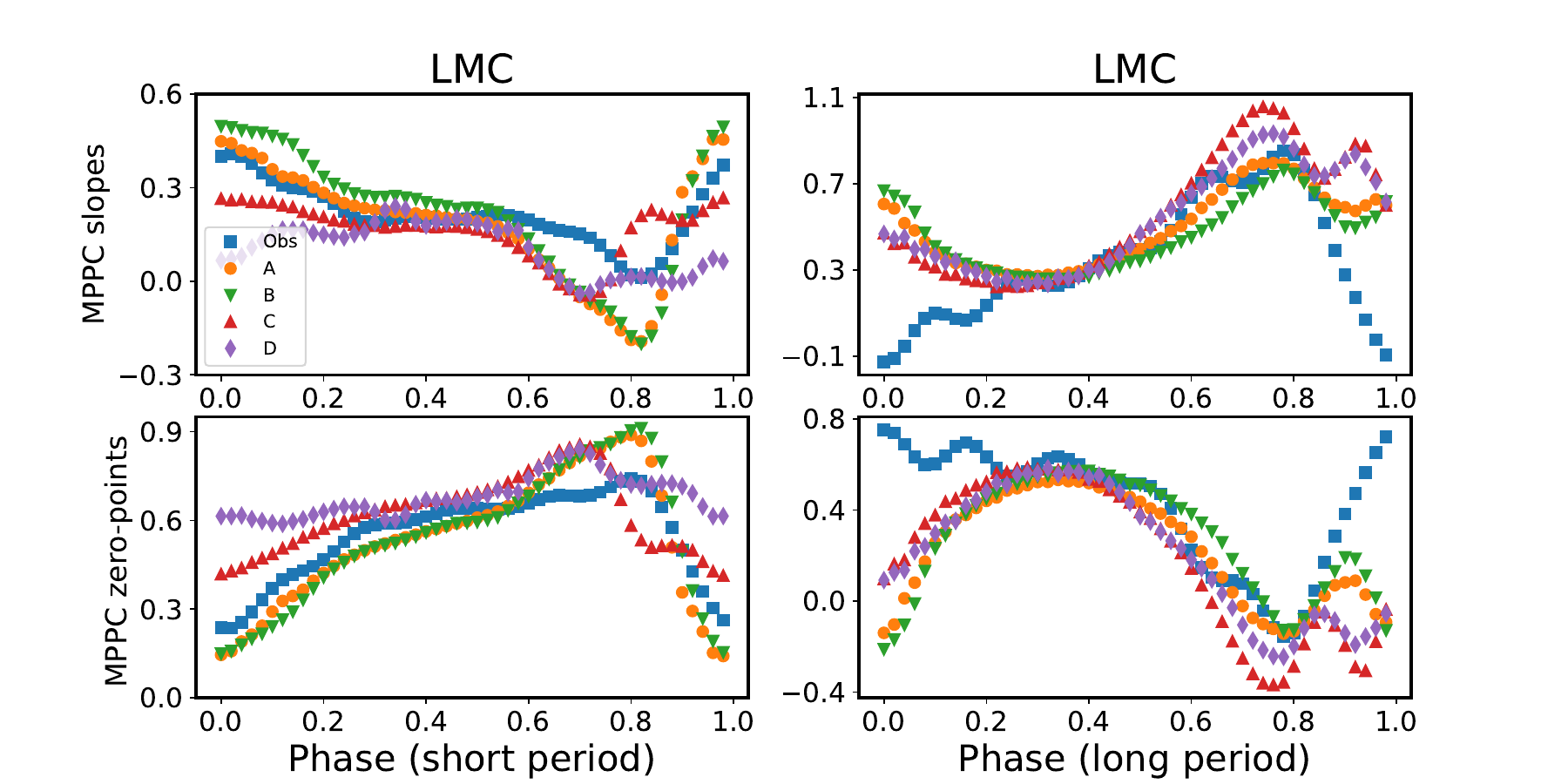}
\caption{Same as Figure~\ref{mppl_lmc} but for MPPC in the LMC.}
\label{mppc_lmc}
\end{figure}

\begin{figure}
\includegraphics[scale=0.5]{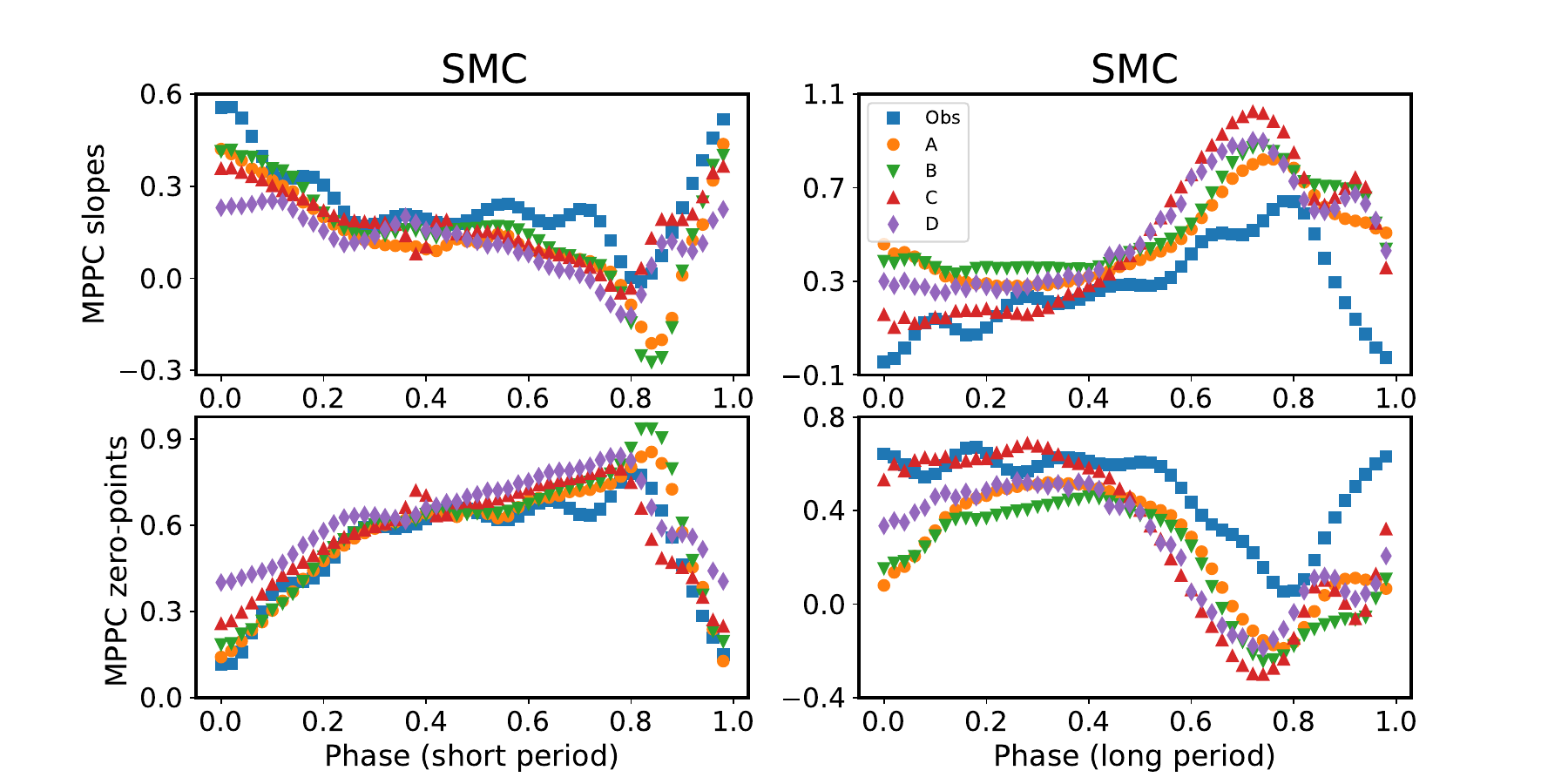}
\caption{Same as Figure~\ref{mppl_lmc} but for MPPC in the SMC.}
\label{mppc_smc}
\end{figure}

The coefficients of the multiphase period-luminosity (MPPL) relations in both $VI$-bands for both the models and observations are displayed in Figures-~\ref{mppl_lmc},~\ref{mppl_lmci},~\ref{mppl_smc},~\ref{mppl_smci}. A distance modulus of $\rm \mu_{LMC}=18.49\pm0.008(\rm statistical)\pm0.047(\rm systematic)$ mag \citep{2019Natur.567..200P} and $\rm \mu_{SMC}=18.97\pm0.016\pm0.028$ mag \citep{2020ApJ...904...13G} are added to the intercept of the theoretical MPPL relations obtained for LMC and SMC, respectively, for comparison with the observed relations. Similarly the coefficients of the multiphase period-color (MPPC) relations from the models and observation are shown in Figures-~\ref{mppc_lmc} and~\ref{mppc_smc}. The results obtained in the study are discussed below:

\begin{enumerate}
\item MPPL Relations: The MPPL slopes/zero-points obtained from the observations display minimum/maximum values at phase $\Phi \sim 0.85$ for short periods for both LMC and SMC Cepheids in both the bands. However, long period Cepheids in $VI$-bands exhibit maximum slopes and minimum zero-points at $\Phi \sim 0.8$ for both LMC and SMC. For LMC, sets A \& B exhibit minimum slopes and maximum zero-points at $\Phi \sim 0.8$, and at  $\Phi \sim 0.7$ using sets C \& D for short periods. In case of the SMC, the same was found at $\Phi \sim 0.85$  using sets A \& B, while those using sets C and D at $\Phi \sim 0.8$. However, for long periods, all the four convection sets exhibit maximum/minimum slopes/zero-points at $\Phi \sim 0.8$ for both LMC and SMC. For both the LMC and SMC, the MPPL relations obtained using convection set A is more consistent with set B, whereas the relations obtained using set C is found to be more consistent with set D. There is a discrepancy between the models and the observations in the MPPL plane. The discrepancy may be because there are fewer models than observations. 

\item MPPC Relations: For both LMC and SMC, the results of MPPC slopes/zero-points using the observations display maximum/minimum points at $\Phi \sim 0.85$ for short periods. However, for long periods, the largest variation of the slopes/zero-points was found at $\Phi \sim 0.8$. At long periods, the LMC models (sets A, B, C, D) exhibit maximum/minimum slopes/zero-points at $\Phi \sim 0.8$ and the SMC models at $\Phi \sim 0.75$. For LMC models at short periods, convection sets A, B display minimum/maximum slopes at $\Phi \sim 0.85$, and sets C \& D at $\Phi \sim 0.7$. In case of the SMC, all the four convection sets display maximum/minimum slopes/zero-points at $\Phi \sim 0.7$. 

\end{enumerate}

\begin{table}[h!]
\caption{Summary of light curves used in the analysis.}
\label{tab:model_number_lmcsmc}
\begin{center}
\begin{tabular}{c c c c c c c} \\ \hline \hline
       & & Set A & Set B & Set C & Set D & Observations \\ \hline
LMC & $N_{s}$    & $169$&   $108$ & $115$  & $66$ &$1407$  \\
    & $N_{l}$    & $138$&   $129$& $142$ & $147$ & $123$ \\ \hline
SMC & $N_{s}$   & $203$ &$149$& $155$ & $87$ & $612$ \\ 
    & $N_{l}$   & $90$ & $85$ & $85$ & $84$ & $91$ \\ \hline
\end{tabular}
\tabnote{$N_{s}$: Short period light curves $N_{l}$: Long period light curves}
\end{center}
\end{table}

\begin{figure}[b]
\includegraphics[scale=0.4]{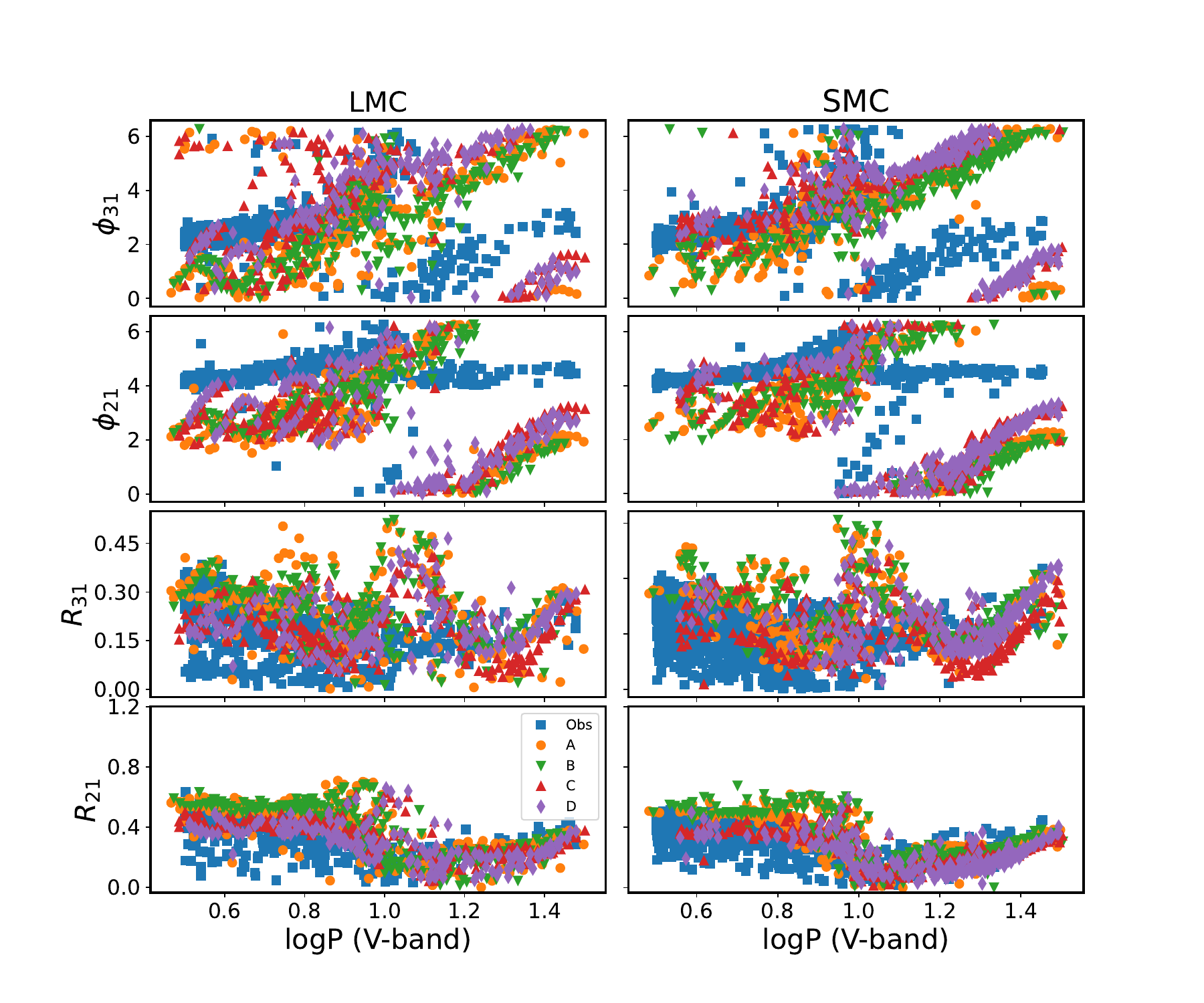}
\caption{Comparison of Fourier parameters obtained using convection sets A, B, C, D with the observed ones for the LMC (left panel) and SMC (right panel) in $V$-band.}
\label{fourier_lmc}
\end{figure}

\section{Discussion and Conclusion}

Analysis of the light curves of Cepheid models generated using \textsc{mesa-rsp} and observations was carried for the LMC and SMC in both $V$- and $I$-bands. The comparison of the results obtained using the models with the observed ones has been done in the multiphase and Fourier parameter planes. The MPPL/MPPC relations obtained from observations and models are found to be highly dynamic in nature, with the effect more pronounced at $\Phi \sim 0.7-0.85$ for short and long periods, in both the LMC and SMC. This is consistent with the results obtained by \citet{2006MNRAS.369..723N}. The significantly smaller number of theoretical light curves compared to observations seem to have a role in the discrepancy between the models and observations in the MPPL plane. The availability of a larger set of non-linear models that cover the whole range of the observation data will provide more details of the apparent discrepancy between the models and the observations.

 In case of the LMC, a discrepancy between the modeled and observed light curves at long periods can be seen at phases between $\Phi \sim 0.0-0.4$ and $\Phi \sim 0.9-1.0$ in the MPPL plane in $VI$- bands, and between $\Phi \sim 0.0-0.2$ and $\Phi \sim 0.9-1.0$ in the MPPC plane. The discrepancy between models and observations at long periods in the MPPL plane is comparatively larger in the SMC compared to the LMC. Similarly, the same can be seen in the MPPC plane. However, from Table-~\ref{tab:model_number_lmcsmc}, it can be seen that the number of long period modeled light curves are comparatively similar that of the observations for both LMC and SMC. From these results, we demonstrate that a thorough investigation of the ensemble of the models, the effects of the $MLTZX$ combinations and the convective parameters on the multiphase relations for long period Cepheids must be considered. 

It is worth mentioning here that the MPPC relations obtained using the four convection sets are found to agree well with the observations for most of the phases in both LMC and SMC. The agreement was observed for both short and long periods Cepheids. 

Interestingly, it has been observed that the nature of the MPPL/MPPC observed relations for short/long period Cepheids exhibit contrasting behaviour for both LMC and SMC, and this behaviour is supported by models as well. This is an interesting result revealed using the multiphase approach and is worthy of a thorough and detailed inspection. 

It can be seen that different convection sets are distinctive from each other in the multiphase plane for most of the phases when compared to the FP plane. Figure~\ref{fourier_lmc} displays the Fourier phase and amplitude parameters as a function of period, obtained using the models and the observations for both LMC and SMC in $V$- band. From the figure, an overlap of the parameters among the four convection sets can be seen. Moreover, the comparison in the multiphase planes requires the ensemble of models must match the ensemble of observations. It may be stated that in this study, interesting results are obtained using the multiphase approach. A deep and detailed investigation considering the shortcomings mentioned in the present study can lead to a better way to test the models and to constrain the pulsation theories.

\section{Acknowledgements}
KK thanks the Council of Scientific and Industrial Research (CSIR), Govt. of India for the Senior Research Fellowship (SRF). KK acknowledges the travel support provided by the International Astronomical Union (IAU) to attend the IAU Symposium 376, “At the cross-roads of astrophysics and cosmology: Period–luminosity relations in the 2020s". SD thanks Council of Scientific and Industrial Research (CSIR), Govt. of India, New Delhi for a financial support through the research grant ``03(1425)/18/EMR-II''. SMK thanks State University of New York, Oswego, NY 23126, USA and Cotton University, Guwahati, Assam for the support. MD thanks CSIR for Junior Research Fellowship provided through CSIR-NET under the project. S. Das acknowledges the support of the KKP-137523 `SeismoLab' Élvonal grant of the Hungarian Research, Development and Innovation Office (NKFIH). AB acknowledges funding from the European Union's Horizon 2020 research and innovation programme under the Marie Skłodowska-Curie grant agreement No.886298. HRR and SK thanks State University of New York, Oswego, NY 23126, USA. The authors acknowledge IUCAA, Pune for the use of High Performance Computing facility Pegasus.

\end{document}